\documentclass[a4paper,10pt]{article}
\usepackage[margin=2cm]{geometry}
\usepackage[english]{babel}
\usepackage[latin1]{inputenc} 
\usepackage[pdftex]{graphicx} 

\newcommand{\dfrac}[2]{\frac{\displaystyle#1}{\displaystyle#2}}
\newcommand{\pder}[2]{\frac{\displaystyle\partial#1}{\displaystyle\partial#2}}

\title{The interaction between the Moon and the solar wind}
\author{M. Holmstr{\"o}m\thanks{Swedish Institute of Space Physics, PO~Box~812, SE-98128~Kiruna, Sweden. (matsh@irf.se)}, S. Fatemi,$^*$ Y. Futaana,$^*$ and H. Nilsson$^*$}
\date{July 8, 2011}

\begin{document}
\maketitle

\begin{abstract}
We study the interaction between the Moon and the solar wind using a 
three-dimensional hybrid plasma solver. 
The proton fluxes and electromagnetical fields are 
presented for typical solar wind conditions with different magnetic 
field directions. 
We find two different wake structures for an interplanetary 
magnetic field that is perpendicular to the solar wind flow, 
and for one that is parallell to the flow. 
The wake for intermediate magnetic field directions will be a 
mix of these two extreme conditions. 
Several features are consistent with a fluid interaction, e.g., 
the presence of a rarefaction cone, and an increased magnetic field in the 
wake.  
There are however several kinetic features of the interaction.  
We find kinks in the magnetic field at the wake boundary. 
There are also density and magnetic field variations in the far wake, 
maybe from an ion beam instability related to the wake refill. 
The results are compared to observations by the WIND spacecraft during 
a wake crossing.  The model magnetic field and ion velocities are in agreement 
with the measurements.  The density and the electron temperature in the 
central wake are not as well captured by the model, probably from the 
lack of electron physics in the hybrid model. 
\end{abstract}

\section{Introduction}
Bodies that lack a significant atmosphere and internal 
magnetic fields, such as the Moon and asteroids, can to a first 
approximation be considered passive absorbers of 
the solar wind. 
The solar wind ions and electrons directly impact the surface of these bodies 
due to the lack of atmosphere, and the interplanetary magnetic field passes 
through the obstacle relatively undisturbed because the bodies are 
assumed to be non-conductive. 
Since the solar wind is absorbed by the body, 
a wake is created behind the object. 
This wake is gradually filled by solar wind plasma downstream of the body, 
through thermal expansion and the resulting ambipolar electric field, 
along the magnetic field lines. 
For a review of the Moon--solar wind interaction, 
including recent findings, see Halekas~{\it et al.} (2010), 
who also point out the fact that we can view the near lunar space as 
a plasma laboratory.  Although the Moon lacks any significant atmosphere 
and ionosphere there are many interesting features to this 
seemingly simple solar wind interaction.  
To understand the global interaction between the solar wind and the 
moon we need computer models that can give us a 
three-dimensional picture that complements the in-situ plasma 
observations.  

Magnetohydrodynamic (MHD) modeling of the Moon--solar wind interaction 
has a long history, starting with the work of Spreiter~{\it et al.}~(1970), 
who analytically treated the special case of an interplanetary magnetic field 
(IMF) aligned with the solar wind flow direction by finding a 
transformation of the MHD equations into a hydrodynamic problem.  
However, there are many kinetic processes that cannot be described by 
fluid models, e.g., the non-Maxwellian particle populations in the 
wake region, so particle models should capture more of the physical 
processes. 
An approximation of the refill of the lunar wake is the general 
one-dimensional problem where plasma expands into a vacuum 
(Widner, 1970; Samir, 1983; Mora, 2003), and such models have specifically been 
applied to the refill of the lunar wake 
(Farrell, 1998; Birch and Chapman, 2001).  
A nice property of such models is that even if they 
are one-dimensional, they can to some degree approximate 
a two-dimensional model of the lunar wake, where time corresponds to distance 
downstream the wake, i.e., the one-dimensional model is applied 
perpendicular to the solar wind flow and convects with the flow. 
A disadvantage of one-dimensional models, e.g., along $x$, is that 
the magnetic field component along the $x$-axis is forced to be constant, 
from the requirement that $\nabla\cdot\mathbf{B}=0$. 
This is a severe restriction for modeling the lunar wake, 
where the field component across the wake can increase. 
Due to the computational complexity, particle in cell (PIC) models, 
that include electrons as particles,  
are for the foreseeable future limited to at most two spatial dimensions, 
so what has been investigated using such models 
is the interaction of an infinite cylinder with the solar wind 
(Harnett and Winglee, 2003; Kimura and Nakagawa, 2008). 

Hybrid models, where ions are particles and electrons a fluid, 
are less computational demanding that full PIC models and are 
suitable for studying processes of size comparable to the 
ion inertial length and gyro radius. 
A two-dimensional hybrid model by 
Tr{\'a}vn{\'{\i}}{\v c}ek~{\it et al.}~(2005) focused mostly on 
wave activity in the lunar wake. 
The Moon--solar wind interaction is however a fully three-dimensional 
problem, since the IMF component perpendicular to the solar wind flow 
introduces asymmetry into an otherwise cylindrical symmetric problem. 
A three dimensional hybrid model by Kallio~(2005)
had a fairly coarse grid on a small region, 
and did not handle the wake region in a self consistent way 
(the electric field was specified in that region).
Recently another three dimensional hybrid model has been presented 
in Wiehle~{\it et al.}~(2011). They compare observations by one of the 
ARTEMIS spacecrafts to a model run with time dependent inflow 
conditions, and find a good agreement. 
The time dependence enables comparison with observations even 
during changing solar wind conditions, but makes it more difficult 
to see how the wake forms differently depending on the 
solar wind conditions. 

Here we also apply a three dimensional hybrid model to the 
Moon--solar wind interaction, but for steady inflow conditions.  
This allow us to study how the wake depends on the direction of the IMF, 
and in what follows we describe the method and 
present a comprehensive global overview of the 
different plasma quantities in the wake.

\section{Model}
In the hybrid approximation, ions are treated as particles, 
and electrons as a massless fluid. 
In what follows we use SI units. 
The trajectory of an ion, $\mathbf{r}(t)$ and $\mathbf{v}(t)$, 
with charge $q$ and mass $m$, is computed from the Lorentz force, 
\[
  \dfrac{d\mathbf{r}}{dt} = \mathbf{v}, \quad
  \dfrac{d\mathbf{v}}{dt} = \dfrac{q}{m} \left( 
    \mathbf{E}+\mathbf{v}\times\mathbf{B} \right), 
\]
where $\mathbf{E}=\mathbf{E}(\mathbf{r},t)$ is the electric field, 
and $\mathbf{B}=\mathbf{B}(\mathbf{r},t)$ 
is the magnetic field.  The electric field is given by 
\begin{equation}
  \mathbf{E} = \dfrac{1}{\rho_I} \left( -\mathbf{J}_I\times\mathbf{B} 
  +\mu_0^{-1}\left(\nabla\times\mathbf{B}\right) \times \mathbf{B} 
  - \nabla p_e, \right) \label{eq:E}
\end{equation}
where $\rho_I$ is the ion charge density, 
$\mathbf{J}_I$ is the ion current, 
$p_e$ is the electron pressure, and
$\mu_0=4\pi\cdot10^{-7}$ [Hm$^{-1}$] is the magnetic constant. 
We assume that the electrons are an ideal gas, 
then $p_e=n_ekT_e$, so the pressure is directly related to temperature 
($k$ is Boltzmann's constant). 
There are several ways to handle the electron 
pressure~(p.\ 8790, Winske and Quest, 1986). 
Here we assume that $p_e$ is adiabatic (small collision frequency), 
with an adiabatic index, $\gamma=5/3$.
Then the relative change in 
electron pressure is related to the relative change in electron density by 
$p_e\propto|\rho_e|^\gamma$~(Winske and Quest, 1986), and we have that 
\[
  \frac{p_e}{p_{e0}} = \left( \frac{n_e}{n_{e0}} \right)^{\gamma}, 
\]
where the zero subscript denote reference values 
(here the solar wind values at the inflow boundary). 
From charge neutrality and $p_e=n_ekT_e$ we can derive that 
\begin{equation}
  p_e = A \rho_I^{\gamma} 
    \mbox{ with } A=\frac{k}{e}\rho_I^{1-\gamma}T_e   \label{eq:pe}
\end{equation}
a constant that is evaluated using reference values of $\rho_I$ and $T_e$ 
(solar wind values). 
Note that $\gamma=1$ corresponds to assuming that $T_e$ is constant, and 
$\gamma=0$ gives a constant $p_e$. 

Finally, Faraday's law is used to advance the magnetic field in time, 
        \[
          \pder{\mathbf{B}}{t} = -\nabla\times\mathbf{E}. 
        \]

We use a cell-centered representation of the magnetic field on a uniform 
grid.  All spatial derivatives are discretized using standard second order 
finite difference stencils.  
Time advancement is done by a predictor-corrector leapfrog method 
with subcycling of the field update, denoted cyclic leapfrog (CL) 
by Matthews~(1994). An advantage of the discretization is that the 
divergence of the magnetic field is zero, down to round off errors. 
Also, the discretization conserves energy very well (Holmstr\"{o}m, 2011). 
The ion macroparticles (each representing a large number of real particles) 
are deposited 
on the grid by a cloud-in-cell method (linear weighting), and 
interpolation of the fields to the particle positions are done by 
the corresponding linear interpolation. 
Initial particle positions are drawn from a uniform distribution, 
and initial particle velocities from a drifting Maxwellian distribution. 
Further details on the hybrid model can be found in Holmstr\"{o}m~(2010). 

The implementation of the algorithm was done in the FLASH software 
framework, developed at 
the University of Chicago (Fryxell~{\it et al.}, 2000), 
that implements a block-structured 
adaptive (or uniform) Cartesian grid and is parallelized using 
the Message-Passing Interface (MPI) library for communication. 
Further details on the FLASH hybrid solver can be found in 
Holmstr\"{o}m~(2011).  The general solver will be included 
in future releases of FLASH. 

A crucial point in modeling the interaction between the Moon and 
the solar wind is how to choose the inner boundary conditions, 
at the surface of the moon. 
Here we model the Moon as an absorber of the solar wind. 
Ideally we should then remove all ions that hit the surface of the Moon, 
but that would introduce a discontinuity in the charge density, 
from solar wind values outside the sphere, to zero inside the sphere. 
However, numerical differentiation of discontinuous functions on a grid 
will lead to oscillations and instabilities in the fields. 
Therefore we choose to keep also ions inside the Moon, but we gradually 
reduce their weight by a factor $f_{obs}$ after each time step while they are 
inside the Moon (the mass and charge of each particle are reduced). 
Thus, the Moon acts as a gradual sink for ions, 
and this approach also avoids the problem of $\rho_I=0$ in 
Eq.~\ref{eq:E}. 
However, we do not plot these reduced ions in what follows, only 
protons that have never hit the Moon are used when computing 
the various statistics. 
This can be seen as a form of smoothing, done in a self consistent way. 
Often smoothing of fields are used in hybrid solvers, but it is then done 
in an ad-hoc way, as an extra step. 
The downside of our approach is that we introduce a numerical parameter, 
but experience shows that the solution will not change by much 
when reducing $f_{obs}$, indicating that the solution is converging. 
The reduction parameter cannot be chosen too small however, 
since a steep gradient in ion density will cause oscillations 
in the magnetic field. 
We have verified numerically that this approach keeps the IMF constant 
inside the Moon, and thus that no spurious currents are present there. 
There are no inner boundary conditions on the electro-magnetic fields. 

\begin{figure}[htbp]
\centerline{\includegraphics[width=0.8\columnwidth, clip]{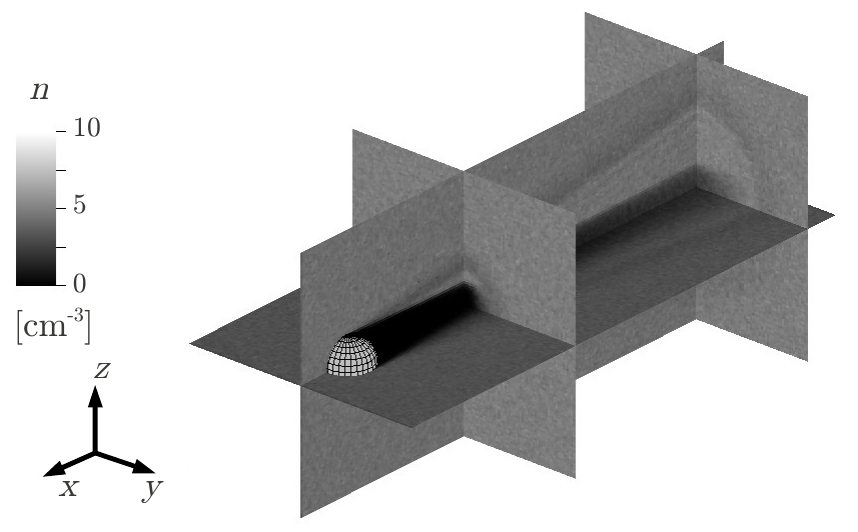}}
\caption{The proton number density for an IMF along 
the direction $(1,1,0)$. 
This also illustrates the coordinate system and the simulation geometry. 
The cuts are the planes $x=-10 000$~km, $x=-30 000$~km, $y=0$, and $z=0$. 
These cutting planes are the same that later are used in Fig.~2-4. 
} \label{fig:geometry}
\end{figure}

The coordinate system is centered at the Moon, with 
the $x$-axis is directed toward the Sun, so the solar wind 
flows opposite to the $x$-axis. 
The IMF is in the $xy$-plane, and we will from now on denote it 
as the IMF plane.  This defines a plane of symmetry, since the 
only asymmetry in the problem stems from a non-zero component 
of the IMF perpendicular to the $x$-axis. 
The solar wind consists only of protons (and the massless electron fluid), 
and we have a rectangular simulation domain divided into cells. 
The $+x$-face of the domain is an inflow boundary where we have a layer of 
shadow cells outside the domain that are filled with macroparticles 
drawn from the solar wind proton velocity distribution, 
and where the solar wind magnetic field is set. 
The $-x$-face is an outflow boundary where 
particles exiting the domain are removed, and where the magnetic 
field is extrapolated from the interior. 
The four other faces have periodic boundary conditions for fields and 
particles. 
The simulation domain is divided into a Cartesian grid with 
cubic cells of size 160~km.  
The geometry of the simulation domain is shown in Fig.~\ref{fig:geometry}. 
The extent of the simulation 
box is $[-32.40,4.08]\times[-8.32,8.32]\times[-8.32,8.32] \cdot 10^3$ km
$\approx [-18,2.4]\times[-4.8,4.8]^2 R_L$, 
with the Moon as a sphere of radius $R_L=1730$~km at the origin.  
The total number of simulation macroparticles is 
about 300 million, with 121 particles per cell on average. 
Here we have used values of $\gamma=5/3$, $f_{obs}=0.965$, 
a time step of $0.08$~s, and a final time of 170~s, when the 
solution has reached a steady state. 
The plots in what follows show the solution at the final time. 
Five subcycles of the CL algorithm are used here when updating 
the magnetic field, 
and the execution time is about 13 hours on 384 CPU cores.

\section{Results}
In what follows, we have used typical solar wind conditions at 1~AU 
(Table 4.1, Kivelson and Russel, 1995), unless otherwise noted. 
The solar wind velocity is 450~km/s, the number density is 7.1~cm$^{-3}$, 
and the ion temperature is $1.2\cdot10^5$~K.  
The electron temperature is $1.4\cdot10^5$~K, and 
the IMF is in the $xy$-plane with a magnitude of 7~nT. 
For these plasma parameters, the ion inertial length is 85~km, 
the Alfv\'{e}n velocity is 58~km/s, 
the thermal proton gyro radius is 66~km (the gyro time is 9~s), 
and the ion plasma beta is 0.6.

\subsection{Overview}
\begin{figure*}[t]
\centerline{\includegraphics[width=0.9\textwidth,clip]{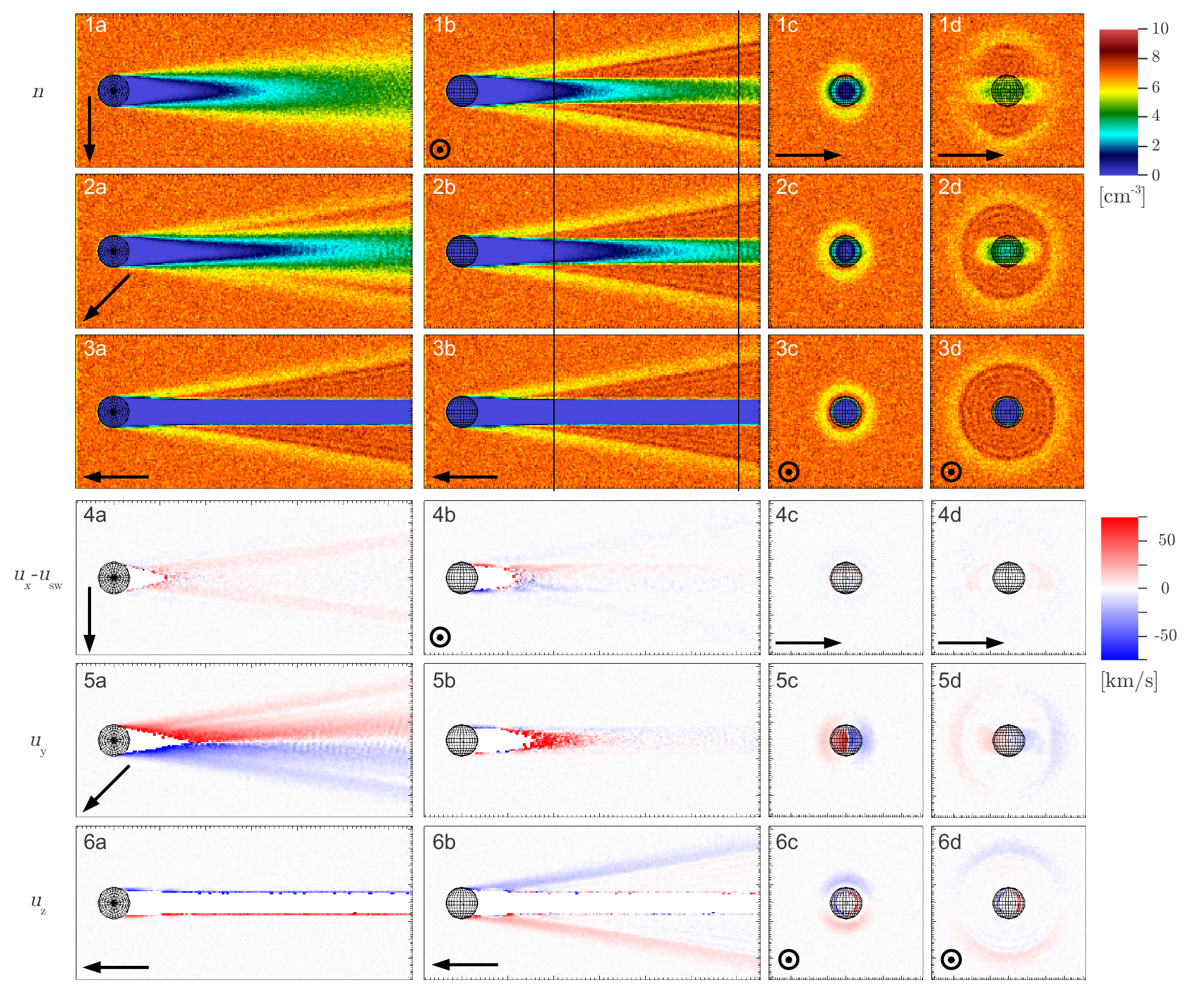}} 
\caption{Proton number density and velocities in different planes, 
for different upstream IMF conditions. 
Arrows show the direction of the IMF. 
Top three rows show number density for different upstream IMF directions. 
Along the directions $(x,y)=(0,1)$, $(1,1)$, and $(1,0)$, respectively. 
The geometry is repeated in the lower three rows (4-6) 
for the proton velocities $u_x-u_{sw}$, $u_y$, and $u_z$. 
The columns are from the left, cuts in the planes $z=0$ (seen from $+z$), 
$y=0$ (seen from $+y$), $x=-10 000$~km, and $x=-30 000$~km (seen from $+x$). 
Fig.~1 provides a three-dimensional illustration of the cutting planes. 
The two vertical lines in 1-3b show the position of the cuts perpendicular to 
the $x$-axis. 
} \label{fig:flow}
\end{figure*}

In Fig.~\ref{fig:flow} the number density (three first rows) 
and the ion velocities (three last rows) are shown, 
each for three different IMF directions 
(at 0, 45 and 90 degrees to the $x$-axis, in the $xy$-plane). 
The most obvious feature 
of the interaction between the Moon and the solar wind is that 
a wake is formed behind the Moon, visible as a region of low density. 
We also see how the refill of the wake depends on the IMF direction. 
The wake refills along the magnetic field lines, making the wake shorter 
when the component of the IMF that is perpendicular to the 
solar wind is larger (Fig.~\ref{fig:flow}.1). 
When the IMF is anti-parallel to the solar wind flow there is 
hardly any refill of the wake over the length of the simulation domain, 
also the flow then is cylindrical symmetric around the Moon--Sun axis. 
That the wake refill occur along the magnetic field lines is also 
visible in the cuts of the far wake (Fig.~\ref{fig:flow}.1d), 
with the refill region as a rectangle aligned with the IMF. 

\begin{figure*}[t]
\centerline{\includegraphics[width=0.9\textwidth,clip]{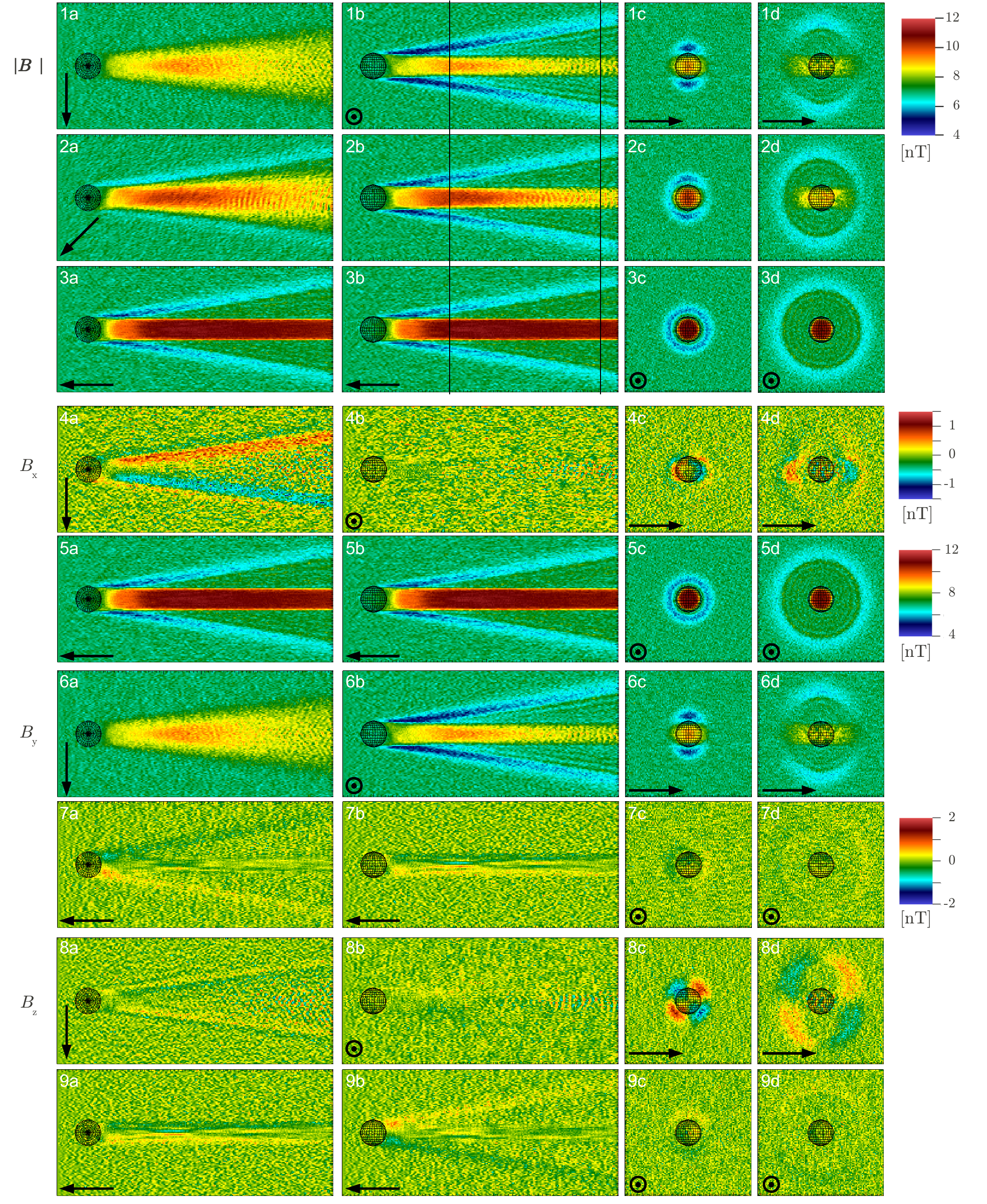}} 
\caption{The magnetic field magnitude, and components in different planes, 
for different upstream IMF conditions. Arrows show the direction of the IMF. 
Top three rows show magnetic field magnitude for different upstream 
IMF directions. Then the magnetic field $x$-component for an IMF 
perpendicular to and opposite the solar wind flow is shown on 
row 4 and 5.  Similarly, the $y$-component is shown on row 6 and 7, 
and the $z$-component on row 8 and 9. 
The geometry of the cuts in the different columns are the same as in Fig.~2. 
The colorbar is for the row immediately to the left, and for all 
following rows, until the next colorbar. 
} \label{fig:bfield}
\end{figure*}

Another notable feature is the density rarefaction cone that emanate 
from the terminator region.  This region of low density propagate mostly 
above the poles in the case of a perpendicular IMF (Fig.~\ref{fig:flow}.1), 
indicating that it propagates perpendicular to the magnetic field lines. 
This will be discussed in more detail when we look at the associated 
reduction in magnetic field strength (Fig.~\ref{fig:bfield}). 

We now examine the ion velocities, computed from the ion current and 
the charge density as $\mathbf{J}_I/\rho_I$, as is usual for particle in cell 
models, since the ion current density and the ion charge density are 
quantities that are available on the computational grid. 
The small white region downstream of the Moon is a region without any 
particles that has not hit the Moon, thus no velocities can be computed 
there. 
The $x$-velocity is shown as the deviation from the solar 
wind velocity, when the IMF is perpendicular to the flow, 
in Fig.~\ref{fig:flow}.4. 
We see a slight slow down of the solar wind in the rarefaction cone. 
There is also a slow down behind the $+z$ hemisphere
(and a speed up behind the $-z$ hemisphere) seen in Fig.~\ref{fig:flow}.4b. 
This is probably a geometric effect, where part of the velocity 
distribution function is emptied by collisions with the lunar surface. 
The $y$-velocity (Fig.~\ref{fig:flow}.5) shows how the wake symmetrically 
refills from both sides of the wake. 
The ions move inward not only in the central part of the wake, 
but also in the rarefaction cone. 
For the case of an aligned IMF, the $z$-velocity (Fig.~\ref{fig:flow}.6) 
also shows a flow in the rarefaction cone toward the wake center. 
There we also see a flow of ions at the wake boundary in the IMF plane, 
and there seem to be a small kinetic effect of alternating  
upward and downward going ions in the upper and lower part of the wake 
(Fig.~\ref{fig:flow}.6b).  This seems associated with the 
wave pattern in density inside the rarefaction cone 
(Fig.~\ref{fig:flow}.1-3), possibly caused by oscillations at the 
wake boundary. 

The magnetic field magnitude (three first rows) is shown for 
three different IMF directions in Fig.~\ref{fig:bfield}, 
where also each of the vector components (rows 4-9) are shown for two different 
IMF directions (perpendicular and aligned to the solar wind flow). 

The most prominent feature is the enhanced magnetic field in the 
wake.  For a perpendicular IMF (Fig.~\ref{fig:bfield}.1) this 
region spreads out in the IMF plane, in the same region 
where we in the density saw the wake refill. 
For the aligned IMF the magnetic field increase is larger, and 
confined to the narrow wake (Fig.~\ref{fig:bfield}.3). 
There are also structured fluctuations in the far wake region in the 
IMF plane.  
Then there is a cone of lower field strength in the case of an 
aligned IMF (Fig.~\ref{fig:bfield}.3), corresponding to the 
cone of reduced density (Fig.~\ref{fig:flow}.3). 
For the perpendicular IMF the cone is partial, and only extends perpendicular 
to the IMF, toward $+z$ and $-z$ (Fig.~\ref{fig:flow}.1). 
Between the central region of enhanced field and the cone of lower field, 
there are field fluctuations parallel to the cone, as is also seen 
in the density (Fig.~\ref{fig:flow}.1b,2b,3b). 
Examining these density plots, the waves seem to originate at the wake 
boundary and could be caused by oscillations at this interface 
between high and low ion density. 

\begin{figure*}[t]
\centerline{\includegraphics[width=0.9\textwidth,clip]{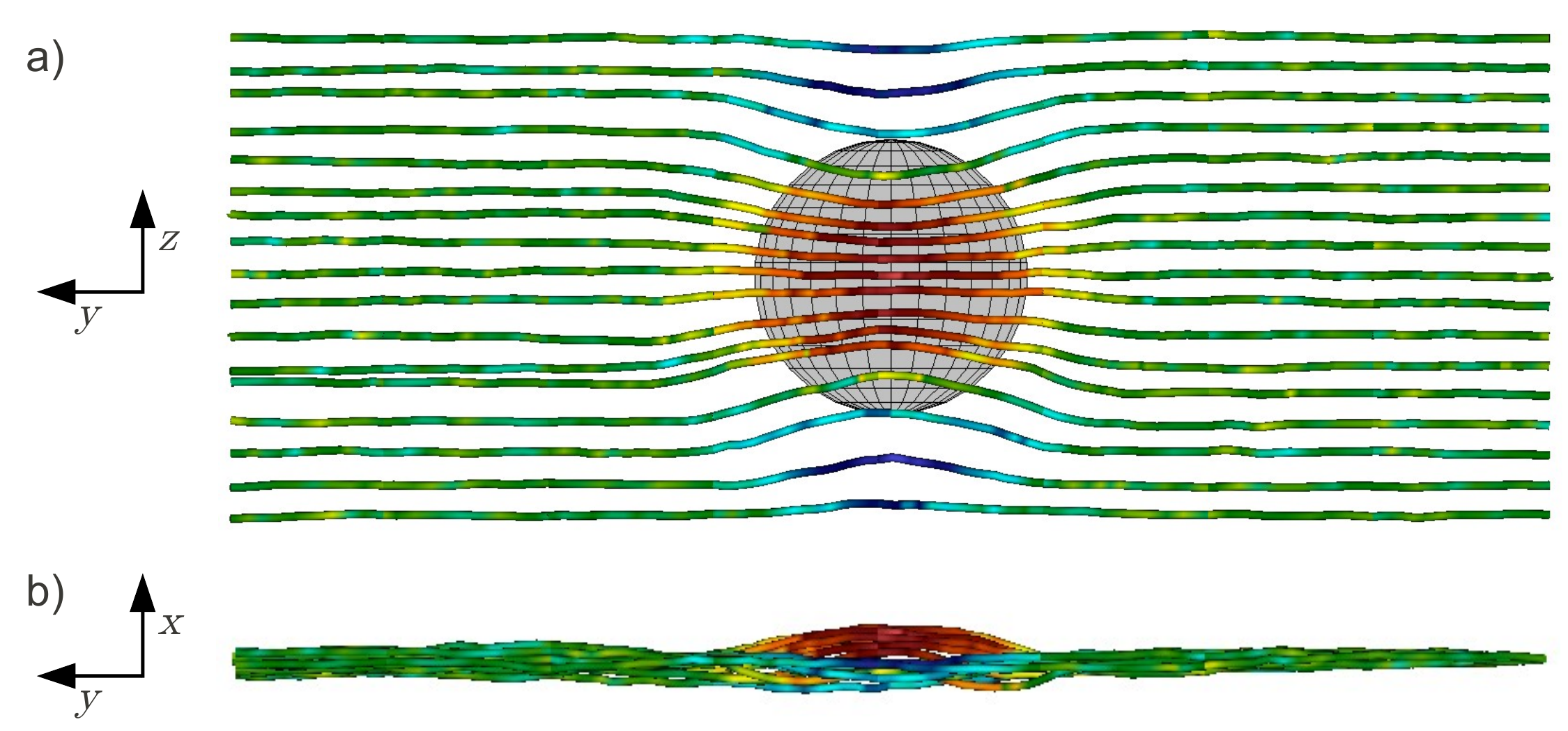}} 
\caption{Magnetic field lines 10000 km behind the Moon for an IMF along 
the $y$-axis (same as in Fig.~\ref{fig:bfield}.1). 
(a) seen along the +$x$-axis, and (b) along the -$z$-axis. 
The color scale shows the magnitude of the magnetic field 
similarly to Fig.~\ref{fig:bfield}.  To enhance the bending of the field the 
magnetic field $x$-component was multiplied by 4, and the $z$-component 
by 2, before drawing the field lines. 
} \label{fig:fieldlines}
\end{figure*}

The magnetic field $z$-component for the case of a perpendicular IMF 
(Fig.~\ref{fig:bfield}.8) exhibits a pinching of the field, 
consistent with earlier predictions (Fig.~3, Owen, 1996) 
and model observations (Kallio,~2005). 
The field is bent toward the IMF plane in the wake 
(Fig.~\ref{fig:bfield}.8c,8d). 
However, the $x$-component of the magnetic field is also perturbed 
(Fig.~\ref{fig:bfield}.4). 
Close to the IMF plane, the field is bent toward the moon on 
one side of the wake, and away from the moon on the other side of 
the wake (Fig.~\ref{fig:bfield}.4a). 
Note also that these bends in the magnetic field are not confined 
to central parts of the wake, but expand along with the rarefaction 
in magnetic field (Fig.~\ref{fig:bfield}.1d) and 
density (Fig.~\ref{fig:flow}.1d). 
Thus, for a perpendicular IMF the magnetic field in the wake is not only 
bent toward the IMF plane, but also toward the moon, as shown in 
Fig.~\ref{fig:fieldlines}. 

The Mach cone of decreased density and magnetic field 
makes an angle to the solar wind 
flow direction of about 10~degrees, as seen in 
Fig.~\ref{fig:flow} and Fig.~\ref{fig:bfield}. 
Since the outer edge of the cone is diffuse, it is impossible to 
give a more exact angle. 
The ion-acoustic wave velocity is 
$v_S\approx \sqrt{2\gamma k_B T_e/m_i}\approx 58$~km/s, and the 
Alfv\'{e}n velocity is also $v_A\approx 58$~km/s for these solar wind 
conditions, corresponding to angles of 7~degrees. 
We see that the density depletion travel perpendicular to the 
magnetic field.  This together with the fact that the propagation 
velocity is larger than both the ion-acoustic and the Alfv\'{e}n velocity 
suggests that the fast magnetosonic wave, with a phase velocity of 
$\sqrt{v_A^2+v_S^2}$ leading to an angle of 10~degrees, 
is responsible for the creation of the Mach cone in the hybrid model. 
This is consistent with the observation of Wiehle~{\it et al.}~(2011) 
that all three MHD modes (fast, Alfv{\'e}n and slow) will be 
manifested in the lunar wake.

\begin{figure*}[t]
\centerline{\includegraphics[width=0.9\textwidth,clip]{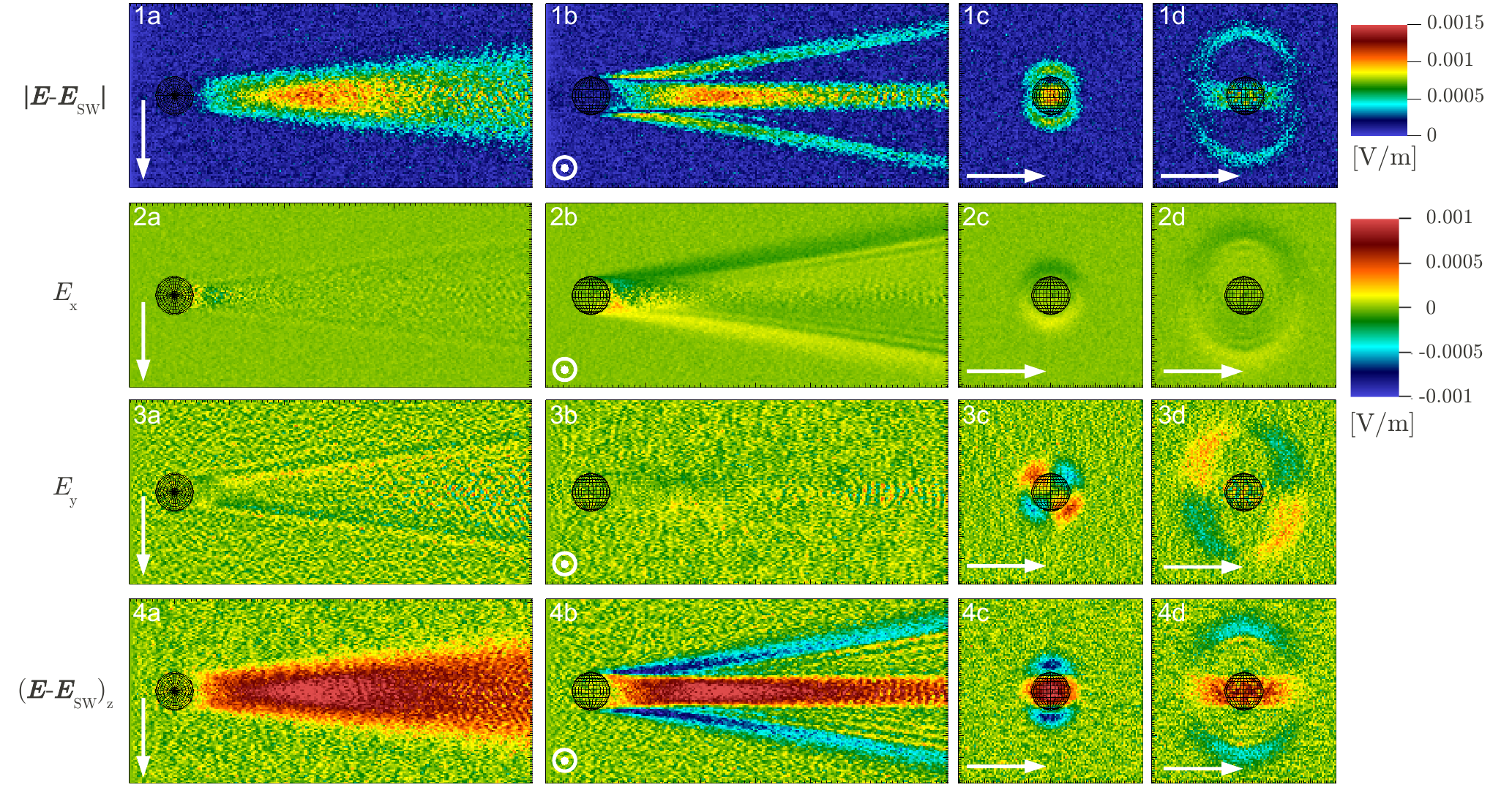}} 
\caption{The electric field magnitude, and the three components 
in different planes, for an IMF that is perpendicular 
to the solar wind flow direction. Arrows show the direction of the IMF. 
For the magnitude and $z$-component, the solar wind convective electric field, 
$\mathbf{E}_{sw}=-u_{sw}\times\mathbf{B}_{sw}$, has been subtracted. 
The cuts in the different columns are the same as in Fig.~2 and~3. 
} \label{fig:efield}
\end{figure*}

In Fig.~\ref{fig:efield} the electric field magnitude (top row) is shown 
along with the three vector components (three lower rows) for the 
case of an IMF perpendicular to the solar wind flow direction. 
What is shown is the electric field with the solar wind 
convective electric field, 
$\mathbf{E}_{sw}=-\mathbf{u}_{sw}\times \mathbf{B}_{sw}$, 
(directed along the $z$-axis) subtracted. 
We do not show the electric field for the case of an IMF aligned with the 
$x$-axis, since it is indistinguishable from zero using the same 
color scale. 
We can note that the regions of enhanced electric field magnitude 
coincide with those of enhanced magnetic field magnitude shown 
in Fig.~\ref{fig:bfield}.1.  Also, there are similarities in the 
components of the electric and magnetic fields.  
$E_y$ is very similar to $B_z$ (Fig.~\ref{fig:bfield}.8), and 
correspondingly $E_x$ is similar to $B_y$ (Fig.~\ref{fig:bfield}.8).

\subsection{Comparison with WIND observations}
We here compare the hybrid model results with observations 
by the WIND spacecraft on December 27, 1994, when it traversed the 
lunar wake approximately 
perpendicular to the solar wind flow, in the plane of the IMF, 
at a distance of $x=-6.5$~R$_L\approx-11300$~km. 
Ogilvie~{\it et al.}~(Fig.~1, 1996) plots several plasma parameters 
during the wake crossing.  Apparently the solar wind conditions 
changed during the observation.  This is most apparent in the 
plasma number density that decreased by more than a factor of two 
from the undisturbed solar wind on the inbound trajectory to the 
solar wind on the outbound trajectory.  
Therefore we make two runs of the hybrid model for the two sets  
of solar wind conditions, here denoted \emph{before} and \emph{after}. 
The solar wind conditions used for the before (after) case are as follows. 
A solar wind velocity of 470 (500)~km/s, 
a number density of 5 (2)~cm$^{-3}$, 
an ion temperature of $0.9$ $(0.5)\cdot10^5$~K, and 
an electron temperature of $1.5$ $(2)\cdot10^5$~K.
The IMF was $(6.39,2.33,0)$~nT for the before case, and 
$(5.57,5.02,0)$~nT for the after case. 
The time step was 0.08 (0.05)~s and $f_{obs}=0.98 (0.985)$.  
Since the plasma parameters are different in the two cases, 
these numerical parameters have different values to ensure 
the stability of the solutions. 

\begin{figure}[htbp]
\centerline{\includegraphics[width=1.0\columnwidth, clip]{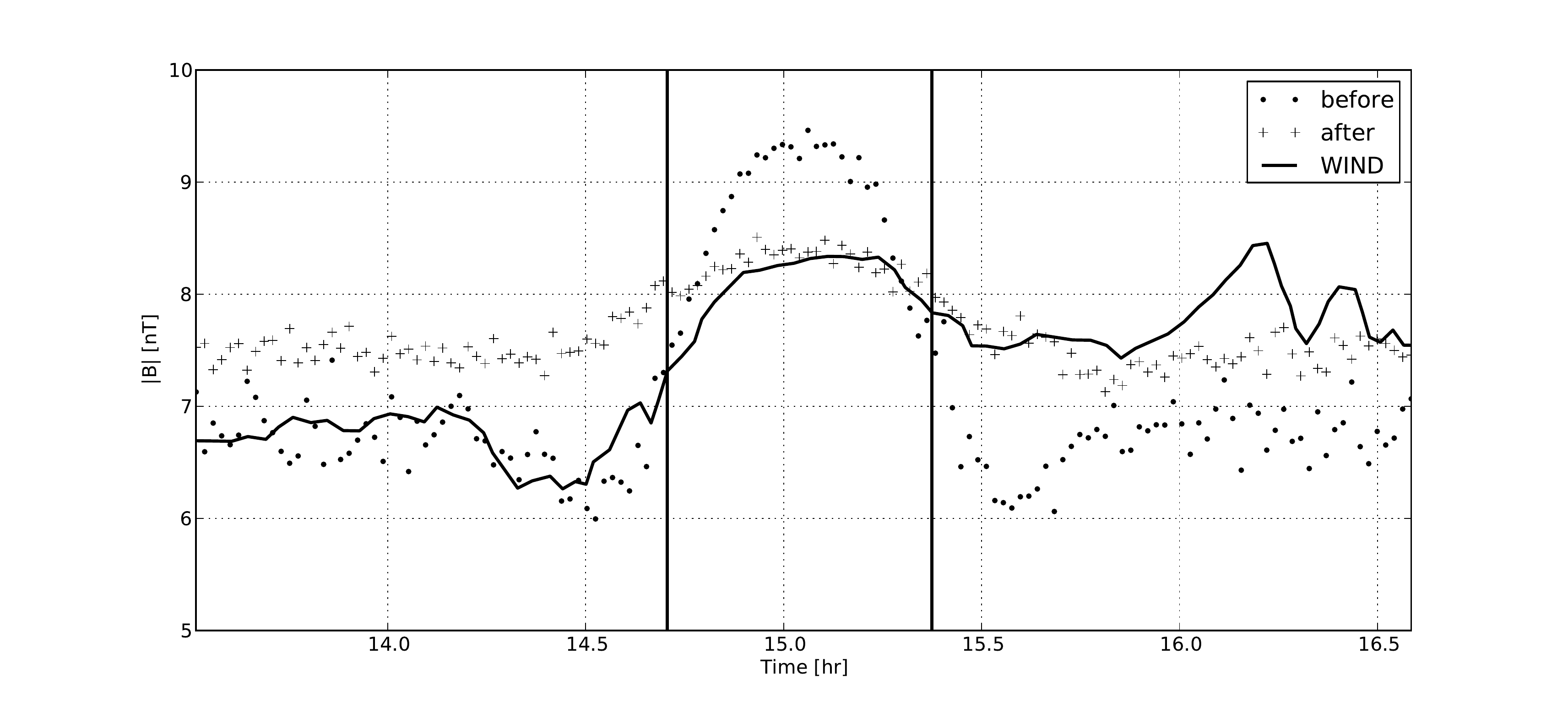}}
\caption{The hybrid model magnetic field magnitude [nT] along 
$y$ at $x=-6.5$ R$_L=$-11245~km for two different upstream 
solar wind conditions (labeled \emph{before} and \emph{after}), 
compared to WIND observations
from Ogilvie~{\it et al.}~(Fig.\ 1, 1996). 
The vertical lines show the location of the optical shadow, 
and the $x$-axis show time on December 27, 1994, as in the original plot. 
} \label{fig:wind_cmp}
\end{figure}

In Fig.~\ref{fig:wind_cmp} we show the model magnetic field magnitudes, 
compared with the observation by WIND. 
The observed magnetic field magnitude matches well to the model values, 
if we assume that there was a change in the upstream solar wind conditions 
around wake entry (there we switch from comparing the \emph{before} 
model results to comparing the \emph{after} model results with the 
observation). 
Before the wake crossing there is a drop in the observed magnetic field 
strength that is also seen in the model. 
This is the rarefaction wave that we saw in Fig.~\ref{fig:bfield} earlier. 
The IMF in the \emph{before} case has the largest component along the 
$x$-axis and therefore the magnetic field magnitude should best match 
Fig.~\ref{fig:bfield}.3a, with a pronounced rarefaction cone. 
Then there is an increase in the central wake which is reproduced in the 
\emph{after} model.  Followed by a gradual decrease at wake exit, 
but no rarefaction wave. 
For the \emph{after} case the IMF is at about 45$^{\circ}$ away from the 
$x$-axis and thus correspond to Fig.~\ref{fig:bfield}.2a, with a much 
weaker rarefaction cone. 
In the observation there are some large variations further away from the 
wake that are not present in the model. 
This could be upstream solar wind disturbances. 

\begin{figure}[htbp]
\centerline{\includegraphics[width=1.0\columnwidth, clip]{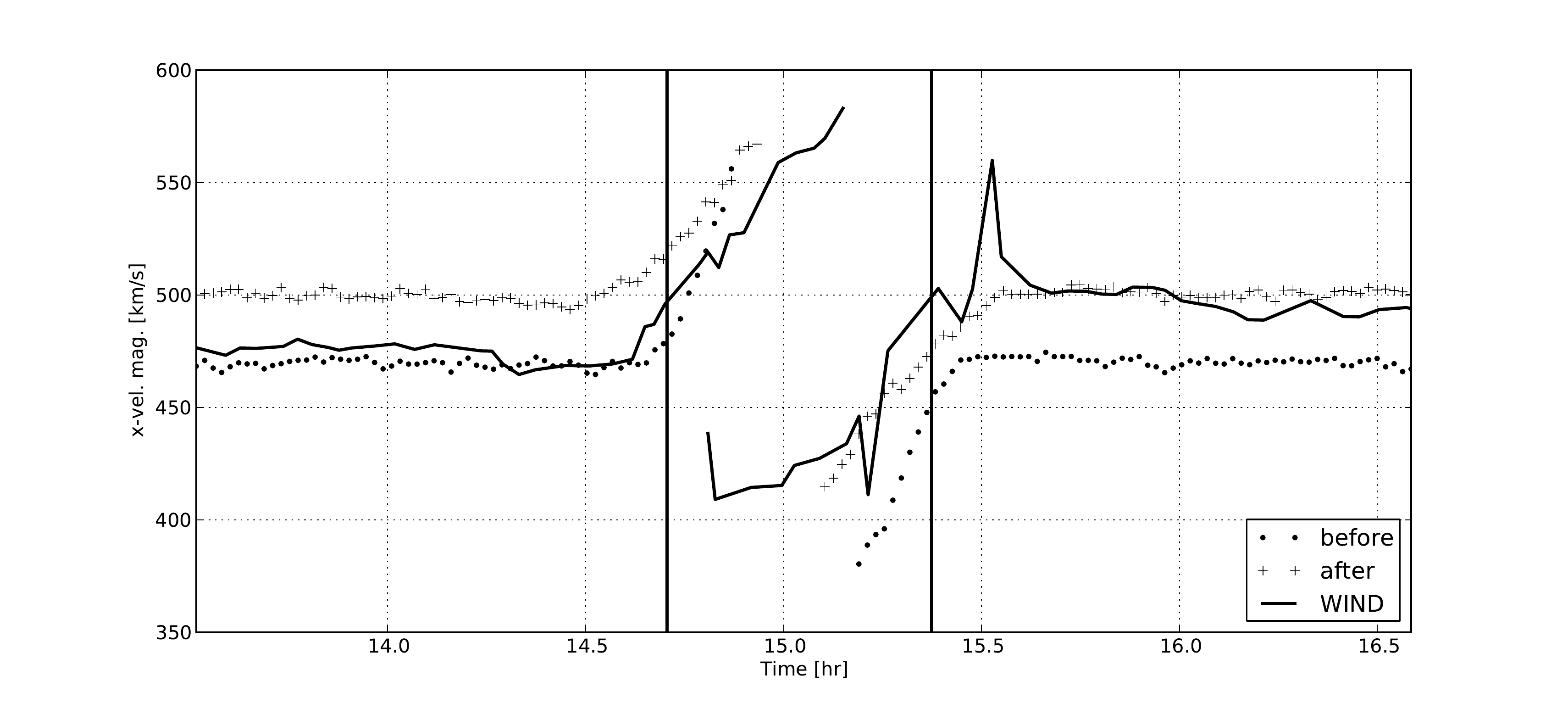}}
\caption{The hybrid model proton $x$-velocity magnitude [km/s]
for two different upstream solar wind conditions
(labeled \emph{before} and \emph{after}), 
compared to WIND observations. 
Otherwise similar to Fig.~\ref{fig:wind_cmp}. 
The reason that the WIND velocity is double valued in the center of the 
wake is that the velocity of the two proton beams refilling the wake 
from each side is computed separately, 
as explained in the original publication (Ogilvie~{\it et al.}, 1996)
} \label{fig:wind_xvel}
\end{figure}

\begin{figure}[htbp]
\centerline{\includegraphics[width=1.0\columnwidth, clip]{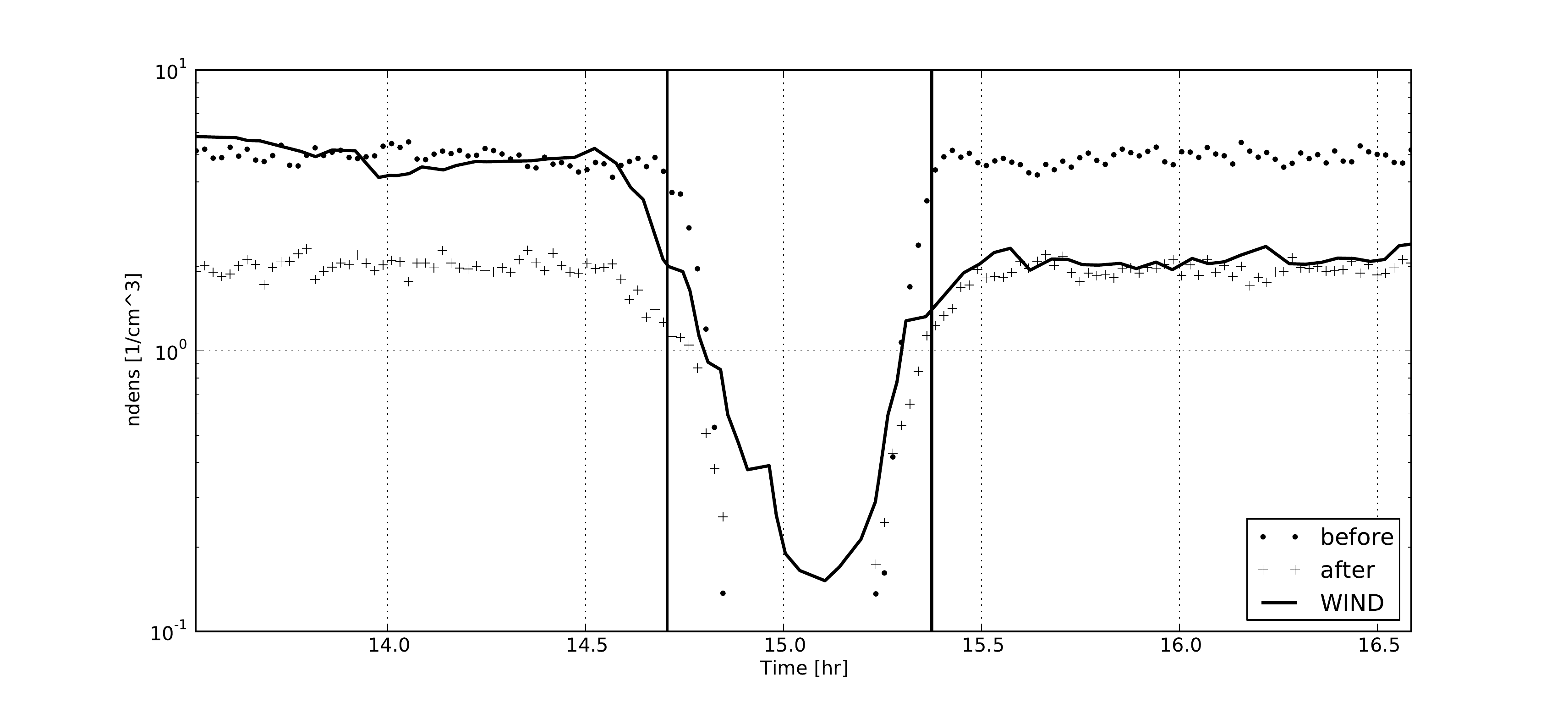}}
\caption{The hybrid model proton number density [cm$^{-3}$]
for two different upstream solar wind conditions, 
compared to WIND observations. 
Otherwise similar to Fig.~\ref{fig:wind_cmp}. 
} \label{fig:wind_ndens}
\end{figure}

In Fig.~\ref{fig:wind_xvel} the proton velocity observed by WIND 
is compared to the velocity in the two hybrid model runs. 
Again we have fairly good agreement between the model and observations, 
assuming a shift in solar wind conditions around wake entry. 
There are more variations in the observed velocities, but the locations 
of the velocity changes, and the slopes match the model. 
In the central wake there is however no data from the model since the 
model density is too low. 
This can be seen when we compare the proton number density from the 
model with observations in Fig.~\ref{fig:wind_ndens}. 
We see that the general features of the observed density profile across the 
wake is found in the model, including a shift of the minimum density 
toward the outbound part of the trajectory, if we assume the shift in 
solar wind conditions around wake entry. 
However, the drop off in density is quicker in the hybrid model, 
and is much below the observations in the central part of the wake. 
The cause of this could be the incomplete treatment of electron dynamics 
in a hybrid model.  Since the electrons are a massless charge 
neutralizing fluid, there are no forces from the electrons on the 
ions, except for the electron pressure gradient term in the 
definition of the electric field (Eq.~\ref{eq:E}). 
In reality, the refill of the wake is governed by the electrons. 
Due to their higher thermal velocity they will speed ahead of the 
ions to refill the wake, setting up an ambipolar electric field 
due to the charge separation that will accelerate the ions 
into the wake (Halekas~{\it et al.}, 2010). 

Similarly, the lack of electron dynamics in the hybrid model will 
affect the electron temperature, $T_e$, in the wake. 
Observations show an increased electron temperature in the wake 
(Fig.~1, Ogilvie~{\it et al.}, 1996). 
But from Eq.~\ref{eq:pe}, and the ideal gas law, 
$T_e \sim \rho_I^{\gamma-1} = \rho_I^{2/3}$ for $\gamma = 5/3$, 
so $T_e$ in the hybrid model is then lower in the wake.

\section{Discussion}
Since the only deviation from axial symmetry is the component of the 
IMF that is perpendicular to the solar wind flow, we can see that 
the two extreme cases are when the IMF is perpendicular to the flow, 
and when the IMF is aligned with the flow. 
For a constant IMF magnitude, the flow and field for cases when 
the IMF is at an angle to the flow will be a mixture of these extreme 
cases, as seen in Fig.~\ref{fig:flow}.2 and Fig.~\ref{fig:bfield}.2.

There are many processes related to electron dynamics that 
are not captured by a hybrid model.  The process of plasma expansion 
into a vacuum has a lot of details, and instabilities that can form 
(Birch and Chapman, 2001).  
Also, the region near the lunar surface 
probably has strong electric fields, and charge separation effects 
(Farrell~{\it et al.}, 2008). 
However using the hybrid approximation it is possible to model the 
global three-dimensional interaction between the Moon and the solar wind, 
so full PIC models and hybrid models are complementary tools in the 
study of this interaction. 
Also, the Moon--solar wind interaction, although seemingly simple, 
is complex enough to be a good test of the applicability of different 
models; fluid, hybrid, and PIC models. 

We have here considered the simplest possible model of the Moon in 
the solar wind --- a spherical sink of solar wind protons. 
There are many physical processes that make the real interaction 
more complex than that.  Two examples are magnetic anomalies that 
deflect the solar wind and will cause downstream disturbances 
(Lue~{\it et al.}, 2011), 
and the reflection of solar wind protons by the dayside lunar surface 
(Holmstr\"{o}m~{\it et al.}, 2010). 

\section{Summary and conclusions}
We have presented a fully self consistent three-dimensional 
hybrid plasma model of the interaction between the Moon and 
the solar wind and shown the resulting global ion fluxes 
and fields for typical solar wind conditions and different IMF directions. 
Magnetic field enhancement in the wake, and a rarefaction wave in 
density and magnetic field are seen in the model. 
The two extreme cases are that of an IMF perpendicular to the solar wind flow, 
and that of an IMF that is parallel to the solar wind flow. 
We find that the wake for intermediate IMF directions can be seen as 
a superposition of these two states. 
Some kinetic effects were also observed in the model.  There are 
density and field variation in the far wake, maybe caused by an ion beam 
instability from the refill of the wake. 
Visible are also kinks in the magnetic field at the inner boundary of the 
rarefaction cone for an IMF 
perpendicular to the solar wind flow direction. 

Observations by the WIND spacecraft were fairly well reproduced by 
the hybrid model, showing that a hybrid model with the moon as 
a spherical sink of ions is able to represent many global 
features of the interaction. 
However, the model central wake density and electron temperature 
were different from observations, probably due to the lack of 
electron dynamics in the hybrid model. 

\subsection*{Acknowledgments}
This work was supported by the Swedish National Space Board, and 
conducted using the resources of the 
High Performance Computing Center North (HPC2N), Ume\aa\ University, Sweden. 
The software used in this work was in part developed by the 
DOE-supported ASC / Alliance Center for Astrophysical 
Thermonuclear Flashes at the University of Chicago.
The work of S.\ Fatemi was supported by the 
National Graduate School of Space Technology,
Lule\aa\ University of Technology.

\subsection*{References}
Birch, P.C., and Chapman, S.C., 
  Detailed structure and dynamics in particle-in-cell simulations
    of the lunar wake, 
  \textit{Physics of Plasmas}, \textbf{8}, 4551--4559, 2001. 

Farrell, W.~M., Kaiser, M.~L., Steinberg, J.~T., and Bale, S.~D.,
  A simple simulation of a plasma void: 
    Applications to Wind observations of the lunar wake, 
  \textit{J. Geophys. Res.}, \textbf{103}, 23653--23660, 1998. 

Farrell, W.~M., Stubbs, T.~J., Halekas, J.~S., Delory, G.~T., 
  Collier, M.~R., Vondrak, R.~R., and Lin, R.~P., 
  Loss of solar wind plasma neutrality and affect on surface potentials 
    near the lunar terminator and shadowed polar regions, 
  \textit{Geophys. Res. Lett.}, \textbf{35}, L05105, 2008.

Fryxell, B., Olson, K., Ricker, P., Timmes, F.~X., Zingale, M., Lamb, D.~Q., 
  MacNeice, P., Rosner, R., Truran, J.~W., and Tufo, H., 
  FLASH: An Adaptive Mesh Hydrodynamics Code for Modeling 
                  Astrophysical Thermonuclear Flashes, 
  \textit{The Astrophysical Journal Supplement Series}, 
  \textbf{131}, 273--334, 2000.
  \texttt{http://flash.uchicago.edu/}

Halekas, J.S., Saito, Y., Delory, G.T., Farrell, W.M., 
  New views of the lunar plasma environment, 
  \textit{Planetary and Space Science}, in press, 
  doi:10.1016/j.pss.2010.08.011

Harnett, E.M., and Winglee, R.M., 
  2.5-D fluid simulations of the solar wind interacting with 
  multiple dipoles on the surface of the Moon, 
  \textit{J. Geophys. Res.}, \textbf{108}, 1088--, 2003. 

Holmstr\"{o}m, M., Hybrid modeling of plasmas, 
  in \textit{Proceedings of ENUMATH 2009}, 
  Edited by Kreiss, G., {\it et al.}, 451 pp, Springer, 2010. 
  arXiv:1010.3291

Holmstr\"{o}m, M., Wieser, M., Barabash, S., and Futaana, Y., 
  Dynamics of solar wind protons reflected by the Moon,
  \textit{J. Geophys. Res.}, \textbf{115}, A06206, 2010.

Holmstr\"{o}m, M., An Energy Conserving Parallel Hybrid Plasma Solver. 
  Submitted to \textit{Proceedings of ASTRONUM-2010}, 2011. 
  arXiv:1010.3291

Kallio, E., Formation of the lunar wake in quasi-neutral hybrid model, 
  \textit{Geophys. Res. Lett.}, \textbf{32}, L06107, 2005.

Kimura, S., and Nakagawa, T., 
  Electromagnetic full particle simulation of the electric field structure 
  around the moon and the lunar wake, 
  \textit{Earth Planets Space}, 
  \textbf{60}, 591--599, 2008. 

Kivelson, M.G., and Russell, C.T, Eds., 
  \textit{Introduction to Space Physics}, 
  Cambridge, 1995. 

Lue, C., Futaana, Y., Barabash, S., Wieser, M., Holmstr\"{o}m, M., 
  Bhardwaj, A., Dhanya, M.B., and Wurz, P., 
  Strong influence of lunar crustal fields on the solar wind flow, 
  \textit{Geophys. Res. Lett.}, \textbf{38}, L03202, 2011. 

Matthews, A.P., 
  Current Advance Method and Cyclic Leapfrog for 2D Multispecies 
  Hybrid Plasma Simulations, 
  \textit{Journal of Computational Physics}, \textbf{112}, 102--116, 1994.

Mora, P., Plasma Expansion into a Vacuum, 
  \textit{Physical Review Letters}, \textbf{90}, 185002, 2003.

Ogilvie, K.W., Steinberg, J.T., Fitzenreiter, R.J., Owen, C.J., 
  Lazarus, A.J., Farrell, W.M., Torbert, R.B., 
  Observations of the lunar plasma wake from the WIND spacecraft 
  on December 27, 1994, 
  \textit{Geophys. Res. Lett.}, \textbf{23}, 1255--1258, 1996.

Owen, C.~J., Lepping, R.~P., Ogilvie, K.~W., Slavin, J.~A., Farrell, W.~M., 
  Byrnes, J.~B., The lunar wake at 6.8 R$_L$: 
  WIND magnetic field observations,
  \textit{Geophys. Res. Lett.}, \textbf{23}, 1263--1266, 1996.

Samir, U., Wright, K.~H., Jr., and Stone, N.~H., 
  The expansion of a plasma into a vacuum: 
  basic phenomena and processes and applications to space plasma physics, 
  \textit{Reviews of Geophysics and Space Physics}, 
  \textbf{21}, 1631--1646, 1983.

Spreiter, J.R., Marsh, M.C., and Summers, A.L., 
  Hydromagnetic aspects of solar wind flow past the Moon, 
  \textit{Cosmic Electrodynamics}, \textbf{1}, 5--50, 
  D.\ Reidel Publishing Company, Dordrecht--Holland, 1970. 

Tr{\'a}vn{\'{\i}}{\v c}ek, P., Hellinger, P., Schriver, D., and Bale, S.~D., 
  Structure of the lunar wake: Two-dimensional global hybrid simulations, 
  \textit{Geophys. Res. Lett.}, \textbf{32}, L06102, 2005.

Widner, M., Alexeff, I., and Jones, W.~D., 
  \textit{Physics of Fluids}, \textbf{14}, 795--796, 1971. 

Wiehle, S., Plaschke, F., Motschmann, U., Glassmeier, K.-H., Auster, H.~U., 
  Angelopoulos, V., Mueller, J., Kriegel, H., Georgescu, E., Halekas, J., 
  Sibeck, D.~G., McFadden, J.~P., 
  First lunar wake passage of ARTEMIS: Discrimination of wake effects 
    and solar wind fluctuations by 3D hybrid simulations, 
  \textit{Planetary and Space Science}, \textbf{59}, 661--671, 2011. 

Winske, D., and Quest, K.B., 
  Electromagnetic Ion Beam Instabilities: 
    Comparison of one- and two-dimensional simulations, 
  \textit{J. Geophys. Res.}, \textbf{91}, 8789--8797, 1986. 
  
\end{document}